\documentclass[preprint]{IEEEtran}

\usepackage{cite}
\usepackage{graphicx}
\usepackage{dcolumn}
\usepackage{bm}
\usepackage{xcolor}
\usepackage[normalem]{ulem} 
\usepackage{hyperref}
\usepackage{amsmath}
\usepackage[switch,mathlines]{lineno}
\usepackage{url}
\usepackage[caption=false,font=footnotesize]{subfig}
\usepackage{fixltx2e}

\begin{document}

\title{Neutron Response of the EJ-254 Boron-Loaded Plastic Scintillator}
\author{Gino Gabella,
	    Bethany L. Goldblum,
	    Thibault A. Laplace,
	    Juan J. Manfredi,
	    Joseph Gordon,
	    Zachary W. Sweger,
	    Edith Bourret
	\thanks{This work was performed under the auspices of the U.S.\ Department of Energy by Lawrence Berkeley National Laboratory under Contract DE-AC02-05CH11231. The project was funded by the U.S.\ Department of Energy, National Nuclear Security Administration, Office of Defense Nuclear Nonproliferation Research and Development (DNN R\&D). This material is based upon work supported in part by the Department of Energy National Nuclear Security Administration through the Nuclear Science and Security Consortium under Award Number DE-NA0003180.}
	\thanks{Gino Gabella, Thibault A. Laplace, Juan J. Manfredi, Joseph Gordon are with the Department of Nuclear Engineering, University of California, Berkeley, CA 94720 USA (email: gino.gabella@berkeley.edu; lapthi@berkeley.edu).}
	\thanks{Bethany L. Goldblum is with the Department of Nuclear Engineering, University of California, Berkeley, CA 94720 USA, and also with the Nuclear Science Division, Lawrence Berkeley National Laboratory, Berkeley, CA, 94720 USA (email: bethany@nuc.berkeley.edu). }
	\thanks{Edith Bourret is with the Materials Sciences Division, Lawrence Berkeley National Laboratory, Berkeley, CA, 94720 USA.}
	\thanks{Zachary W. Sweger was with the Department of Nuclear Engineering, University of California, Berkeley, CA 94720 USA. He is now with the Department of Physics and Astronomy, University of California, Davis, CA 95616 USA.}
		
}

\maketitle
\begin{abstract}

Organic scintillators doped with capture agents provide a detectable signal for neutrons over a broad energy range. This work characterizes the fast and slow neutron response of EJ-254, an organic plastic scintillator with 5\% natural boron loading by weight. For fast neutrons, the primary mechanism for light generation in organic scintillators is n-p elastic scattering. To study the fast neutron response, the proton light yield of EJ-254 was measured at the 88-Inch Cyclotron at Lawrence Berkeley National Laboratory. Using a broad-spectrum neutron source and a double time-of-flight technique, the EJ-254 proton light yield was obtained over the energy range of approximately 270 keV to 4.5 MeV and determined to be in agreement with other plastic scintillators comprised of the same polymer base. To isolate the slow neutron response, an AmBe source with polyethylene moderator was made incident on the EJ-254 scintillator surrounded by an array of EJ-309 observation detectors. Events in the EJ-254 target coincident with the signature 477.6~keV $\gamma$~ray (resulting from deexcitation of the residual $^{7}$Li nucleus following boron neutron capture) were identified. Pulse shape discrimination was used to evaluate the temporal differences in the response of EJ-254 scintillation signals arising from $\gamma$-ray and fast/slow neutron interactions. Clear separation between $\gamma$-ray and fast neutrons signals was not achieved and the neutron capture feature was observed to overlap both the $\gamma$-ray and fast neutron bands. Taking into account the electron light nonproportionality, the neutron-capture light yield in EJ-254 was determined to be 89.4$\pm$1.1~keVee.
\end{abstract}


\section{Introduction}

Nuclei that exhibit a high neutron capture cross section for example, $^{10}$B and $^{6}$Li, can be loaded in organic scintillators to enable detection of slow neutrons \cite{Duckworth1950, Drake1986, Pawelczak2014, Yemam2015, ZaitsevaIAEA}. This capability is useful for basic nuclear physics studies and has a range of applications in proliferation detection, radiation safety, homeland security, and neutron capture therapy \cite{Yen2000, Rasolonjatovo2002, Ishikawa2004, Swiderski2010, Pino2014}. For example, the DarkSide-50 experiment employs a boron-loaded liquid scintillator for neutron background suppression in the search for dark matter \cite{Agnes2016, Westerdale2017}. In border monitoring applications, organic scintillators doped with capture agents increase the detection efficiency for shielded sources of special nuclear material \cite{Peerani, Swiderski2008}. For antineutrino-based approaches to nuclear reactor monitoring, capture-doped organic scintillators can improve neutron background rejection criteria \cite{Dazeley2018}. 

This work evaluates the fast and slow neutron response of EJ-254, a commercially available boron-loaded organic scintillator from Eljen Technology. The standard EJ-254 formulation is comprised of a polyvinyltoluene (PVT) polymer matrix loaded with approximately 1\% $^{10}$B (for a 5\% natural boron loading by weight) \cite{EJ-254}. This scintillator is useful for fast neutron spectroscopy and provides a detectable signal for thermal neutrons via scintillation light generated primarily by the residual $\alpha$ particle produced in the boron neutron-capture reaction~\cite{Pino2014, Greenwood1979}. 

For fast neutrons (above approximately 1~keV), interactions via n-p elastic scattering give rise to recoil protons that excite the surrounding medium, resulting in the production of scintillation light \cite{Klein2007}. The proton light yield relates the light output of the scintillator to the energy of the recoiling proton and is a unique characteristic of the scintillating material. Knowledge of the proton light yield is necessary for modeling the efficiencies of fast neutron detectors as well as for kinematic neutron image reconstruction \cite{Westerdale2017, Aoyama1993, Weinfurther2018}. 

For slow neutrons (below approximately 1 keV), the number of photons produced by recoil protons becomes negligibly small. Neutron moderation via downscattering in the scintillating volume results in an increased probability of capture on $^{10}$B: 
\begin{align*}
&^{10}\mathrm{B}+n \rightarrow \\
&\begin{cases}	 
^{7}\mathrm{Li}\ + \text{ } ^{4}\mathrm{He} & \text{Q = 2.792 MeV, 6\%} \\ 
^{7}\mathrm{Li}\ + \text{ } ^{4}\mathrm{He}\ +\ \gamma(\text{477.6~keV}) & \text{Q = 2.310 MeV, 94\%.}  \label{eqn}
\end{cases}
\end{align*}
The $^{7}$Li and $^{4}$He nuclei produced in the reaction transfer energy to the system via Coulombic interactions, and the scintillation response is dominated by $\alpha$ excitation due to the ionization quenching effect. For neutrons with energies $<100$~keV, the residual $^{7}$Li nucleus is populated in its first excited state with a 94\% branching ratio \cite{Brown2018}, which decays by prompt $\gamma$ emission with $E_{\gamma} =  477.6$~keV. For faster neutrons, the probability of populating the first excited state in $^{7}$Li decreases with increasing neutron energy. This work uses the characteristic $\gamma$ ray from $^{7}$Li deexcitation to aid in identifying the slow neutron response.

Section \ref{sec:ply} details the experimental setup, data analysis, and results of a measurement of the EJ-254 proton light yield from approximately 270~keV to 4.5~MeV. Section \ref{sec:thermal} describes the experimental setup, data analysis, and results of a measurement of the slow neutron response. Concluding remarks are provided in Section \ref{sec:sum}.

\section{Fast Neutron Response}
\label{sec:ply}

\subsection{Experimental Methods}

The double time-of-flight (TOF) method of Brown et al.\ was used to characterize the relative proton light yield of EJ-254 \cite{Brown18}. A high-flux, broad-spectrum neutron beam was produced by impinging 16 MeV deuterons from the 88-Inch Cyclotron at Lawrence Berkeley National Laboratory onto a 3-mm-thick Be target located in the cyclotron vault~\cite{Harrig18}. The EJ-254 organic plastic scintillator, a 5.08 cm diameter $\times$ 5.08 cm height right circular cylinder mounted on a Hamamatsu H1949-51 photomultiplier tube (PMT) with BC-630 silicone optical grease, was positioned in-beam in an experimental chamber approximately 6.96~m from the Be target. The PMT was biased to $-1650$~V. The EJ-254 scintillator was wrapped in at least ten layers of polytetrafluoroethylene tape to maximize light collection \cite{Janecek2012}.  

\begin{figure}
 	\centering
 	\includegraphics[width=0.5\textwidth]{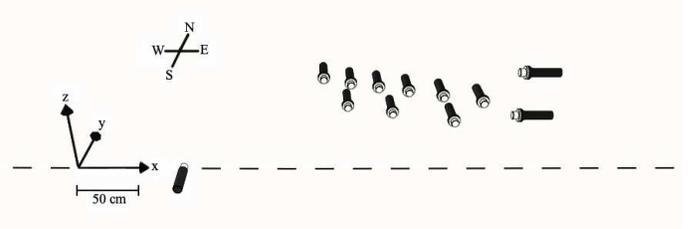}
 	\caption{Experimental setup for the proton light yield measurement. The neutron beam travels along the $x$-axis (illustrated by the dashed line) through the EJ-254 target scintillator. The 11 observation cells were positioned at forward angles with respect to the incoming neutron beam.}
 	\label{PLY-expschem}
 \end{figure}

An array of 11 EJ-309 observation scintillators was placed out of beam and at forward angles to detect n-p elastically scattered neutrons. The EJ-309 organic liquid scintillator from Eljen Technology has pulse shape discrimination (PSD) properties, enabling differentiation of neutron and $\gamma$-ray interactions on an event-by-event basis. Each observation detector consisted of a 5.08 cm diameter $\times$ 5.08 cm height right circular cylindrical aluminum cell filled with EJ-309 with a single borosilicate glass window. Each cell was coupled to the corresponding PMT with BC-630 silicone optical grease. All PMTs were Hamamatsu Photonics Type 1949-50 or 1949-51 and were negatively biased using either a CAEN R1470ET or CAEN NDT1470 power supply. A schematic of the experimental setup is provided in Fig.~\ref{PLY-expschem}.

The energy of the scattered neutron, $E_{n}^{'}$, was determined using the TOF between the target scintillator and the corresponding observation detector. The energy of the recoiling proton, $E_p$, was then calculated on an event-by-event basis, using the neutron scattering angle, $\theta$ (known from the fixed positions of the observation detectors), and the scattered neutron energy: 
\begin{equation}
\label{eq:ep}
E_{p} = E_{n}^{'}\tan^{2}\theta.
\end{equation}
The light output in the target scintillator associated with a given proton recoil energy was determined to provide the proton light yield relation. The TOF of the incoming neutron (determined using the time difference between an event in the target scintillator and the cyclotron RF signal) was used to reduce background contributions~\cite{Brown18}. This kinematically over-constrained system provides a strong rejection criterion for multiple scattering events in the target scintillator \cite{Manfredi2020}. 

\subsection{Data Acquisition and Pulse Processing}
\label{daq-dpp}

Data acquisition was accomplished using a CAEN V1730 500 MS/s digitizer, programmed to record events when a signal was observed in the target detector and at least one of the observation detectors within a 400~ns window. Full waveforms were recorded in list mode with global timestamps. The CAEN digital constant fraction discrimination algorithm (75\% fraction and 4 ns delay) was applied to determine the arrival time of the scintillation pulses and leading edge discrimination was applied for the cyclotron RF timing determination. Data reduction was accomplished using a custom, object-oriented C\texttt{++} library, which used elements of the ROOT v6.16 data analysis framework \cite{Brun1997}. For target pulse processing, a mean waveform was obtained by averaging pulses with energies spanning $25-75$\% of the digitizer dynamic range. The waveform integration length for EJ-254 was set to 300~ns to ensure collection of over 95\% of the scintillation light in the average waveform emitted within a 606~ns acquisition window. Using a pulsed LED circuit adapted from the method of Friend et al.\cite{Friend11}, the target PMT response was shown to be linear within 1\% over the relevant output range.

\subsection{Calibration and Data Analysis}
\label{LYcalib-Analysis}

The light output of the target detector was calibrated using the Compton edge (at 477~keV) of the 662~keV $\gamma$ ray from a $^{137}$Cs source, placed $>5$~cm away from the detector. The light output unit was defined relative to the light yield of a 477~keV electron, that is, a relative light unit of 1 corresponds to the same amount of light produced by a 477~keV recoil electron~\cite{Laplace2020}. Electron-equivalent light units were not used because they are poorly defined if the light output is not proportional to the electron energy. This consideration is relevant to the present work as EJ-254 shares the same PVT polymer base as the EJ-200 plastic scintillator from Eljen Technology and its commercial equivalent BC-408, which are known to have a nonproportional electron response \cite{Nassalski08,Payne11}. A Geant4~\cite{geant4} simulation of the electron energy deposition spectrum convolved with a detector resolution function \cite{dietze} was used to obtain the light output associated with the Compton edge by minimizing $\chi^2$ between the measured and simulated spectra. The parameter minimizations were performed using the SIMPLEX and MIGRAD algorithms from the ROOT Minuit2 package \cite{Brun97}.

Timing calibrations were performed for both the incoming and outgoing neutron TOF measurements. In the case of the incoming neutrons, the TOF was calibrated using the photon flash produced by deuterons impinging on the Be target and the known distance between the Be target and the EJ-254 scintillator. Assuming normality, the standard deviation of this distribution was determined to be 6.97~ns. The outgoing neutron TOF was calibrated individually for each observation detector using time differences between $\gamma$-ray interactions in the target and observation detectors. The standard deviations of the resulting distributions ranged between 0.4 and 0.5~ns. 

\begin{figure}
\centering
\includegraphics[width=0.5\textwidth]{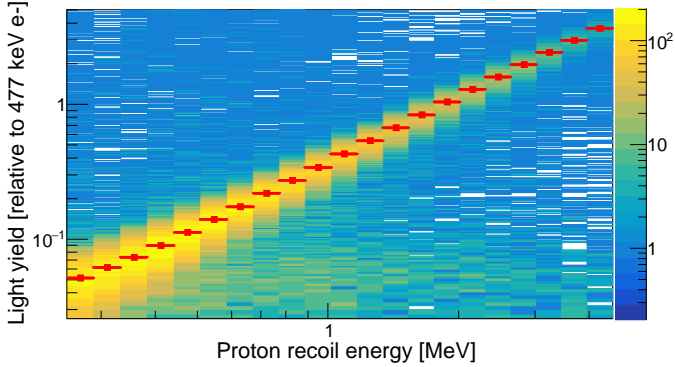}		
\caption{(Color online) 2-D histogram of the EJ-254 relative light yield as a function of proton recoil energy. The binning on the abscissa was set using the recoil proton energy resolution. The red data points indicate the centroids obtained from fitting the projected pulse integral spectra in each bin. The $y$-error bars, which are smaller than the data points in some cases, represent both the statistical and systematic uncertainty.}
	\label{2Dbinned}
\end{figure}

\begin{figure}
\includegraphics[width=0.5\textwidth]{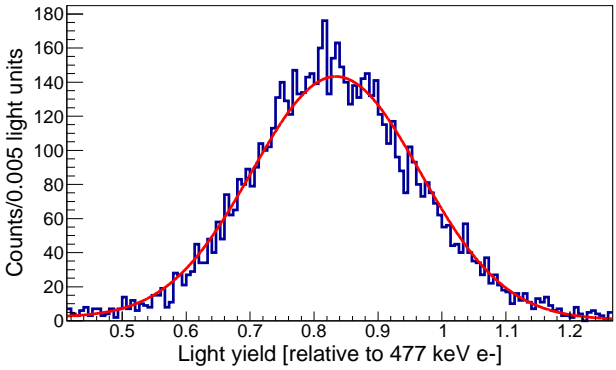}
\caption{(Color online) Relative light yield spectrum (blue) for the proton energy bin centered at $E_p = 1645$~keV fit with a piecewise power law (predominant at low amounts of light) and a normal distribution (red). The centroid of the distribution corresponds to the mean light output for n-p elastic scattering events within the bin. \label{slice}}
\end{figure}
  
Fig.~\ref{2Dbinned} provides a 2-D histogram of the measured EJ-254 relative light yield as a function of proton recoil energy. A series of constraints were applied to the data, including PSD in the observation detectors to isolate neutron events as well as maximum and minimum outgoing neutron energies reflective of the deuteron beam energy and intersection of the n-p elastic scattering feature with the detection threshold. The difference between the measured and kinematically-calculated incoming TOF (the latter obtained using the outgoing TOF and scattering angle) was also required to be less than 15.8~ns (i.e., $<10\%$ of the cyclotron period). The proton energy resolution, determined via a Monte Carlo simulation, was used to define the proton-energy binning in Fig.~\ref{2Dbinned}. A synthetic dataset was generated taking into account the angular uncertainty due to the geometry of the detector array and the uncertainties in the neutron flight path and outgoing TOF. For each simulated neutron-scattering event, the proton recoil energy was calculated using Eq.~\ref{eq:ep} (assuming interaction at the center of the detection volume) and compared to the simulated proton energy to provide the heteroskedastic proton energy resolution function. To reduce the 2D distribution into data points, each bin was projected onto the light-yield axis and fit via maximum-likelihood estimation using a Gaussian function with the background modeled using a continuous piecewise power law distribution. This is illustrated for the proton energy bin centered at $E_p = 1645$~keV in Fig.~\ref{slice}. The red data points in Fig.~\ref{2Dbinned} indicate the centroids obtained from fitting the projected pulse integral spectra in each bin.

The main contributions to the light yield uncertainty were the sensitivity to parameters in the light yield analysis and the uncertainty in the determination of the Compton edge position during light output calibration. The stability of the PMT gain as a function of time was investigated as a potential source of uncertainty and determined to be negligible. To quantify the uncertainty from sensitivity to the light yield parameters, a Monte Carlo calculation was performed in which each parameter was varied by sampling from a distribution defined by its associated uncertainty. The data reduction was then repeated with these varied parameters to determine the influence on the extracted light yield. The sampled parameters were the incoming/outgoing neutron TOF calibration constants, the measured detector locations, and the flight path from the Be target to the origin of the coordinate system. The  uncertainty associated with determination of the Compton edge position as obtained using the error on the fit parameters was negligibly small. As a proxy for model error, the Compton-edge fit was repeated while the fit region about the edge was varied, which resulted in a standard deviation of 1\%. The resulting uncertainties were added in quadrature to arrive at the total light yield uncertainty for each data point. 

 \subsection{Results and Discussion}

 \begin{figure}
 	\centering
 	\includegraphics[width=0.5\textwidth]{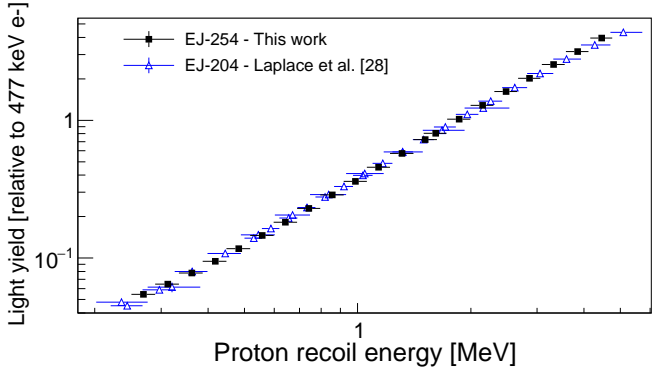}
	\caption{(Color online) EJ-254 proton light yield relation (filled black squares) compared to that of EJ-204 (open blue triangles) by Laplace et al.~\cite{Laplace2020}. The $x$-error bars represent proton-energy bin widths. The $y$-error bars represent the statistical and systematic light output uncertainty. In some cases, the error bars are smaller than the data points.}
 	\label{EJ254andEJ204}
 \end{figure}
 
Fig.~\ref{EJ254andEJ204} shows the relative proton light yield of EJ-254 obtained in this work alongside the EJ-204 proton light yield from Laplace et al.~\cite{Laplace2020}. The error bars on the abscissa represent a bin width, and those on the ordinate axis represent the combined statistical and systematic uncertainty. Since the two plastic scintillators share the same PVT base, agreement is expected between their relative proton light yield data as ionization quenching is a primary process at low fluor concentration \cite{birks}. Indeed, over the full range of the measurement, the EJ-254 and EJ-204 proton light yield relations agree within two standard deviations. The EJ-254 relative proton light yield data are summarized in the Appendix in Table~\ref{resultsTable254}. 

In Fig.~\ref{EJ254EJ276EJ309}, the relative proton light yield of EJ-254 is compared to previous measurements by Laplace et al.~\cite{LaplaceJINST} of the EJ-309 liquid and EJ-276 plastic PSD-capable organic scintillators from Eljen Technology. The error bars on the abscissa represent a bin width, and those on the ordinate axis represent the combined statistical and systematic uncertainty. At 1~MeV, the proton light yield of EJ-254 is approximately 19\% lower than that of EJ-309 and approximately 21\% higher than that of EJ-276. The relatively higher electron number density of the plastic scintillators contributes to higher specific energy loss of the recoil protons, which in turn contributes to increased ionization quenching in the plastic media compared to the liquid EJ-309 scintillator \cite{Birks1951}. As the solvent-to-solute transfer is less efficient in plastic scintillators relative to liquids, higher fluor concentration is typically required to achieve similar scintillation efficiency~\cite{birks321}. For PSD-capable plastic scintillators such as EJ-276, fluor concentrations must be even higher to provide conditions suitable for sufficient triplet-triplet annihilation~\cite{Zaitseva2012}. Concentration quenching, which arises from the formation of metastable dimers of excited and unexcited molecules, increases with increasing fluor concentration. This excimer production would be exacerbated at high excitation density, which may be partly responsible for the relatively lower proton light yield of EJ-276 to EJ-254 to EJ-309.

\begin{figure}
	\centering
	\includegraphics[width=0.5\textwidth]{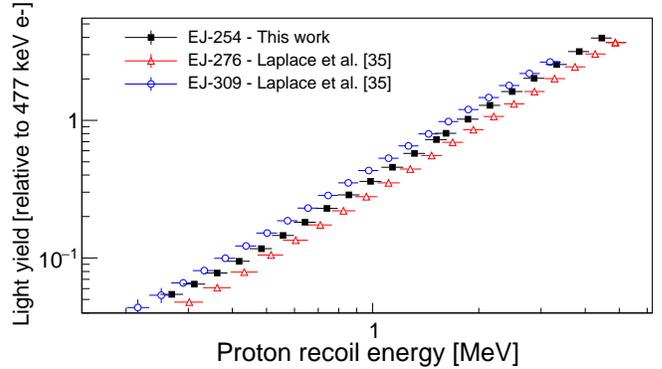}
     \caption{(Color online) EJ-254 proton light yield relation (filled black squares) compared to that of EJ-276 (open red triangles) and EJ-309 (open blue circles) by Laplace et al.~\cite{LaplaceJINST}. The $x$-error bars represent proton-energy bin widths. The $y$-error bars represent the statistical and systematic light output uncertainty. In some cases, the error bars are smaller than the data points.}
	\label{EJ254EJ276EJ309}
\end{figure}

 \section{Slow Neutron Response}
 \label{sec:thermal}
  
 \subsection{Experimental Methods}
  
The identification of neutron-capture events was accomplished using a method adapted from Sun et al.~\cite{Sun2019}. The 5.08 cm dia.\ $\times$ 5.08 cm h.\ right circular cylindrical EJ-254 target scintillator was placed near a 56~mCi AmBe source with nine EJ-309 observation detectors placed at forward angles with flight paths ranging from $0.7$ to $1.2$~m. The target scintillator was  wrapped in at least ten layers of polytetrafluoroethylene tape and optically coupled using BC-630 silicone grease to a Hamamatsu H1949-51 PMT biased to $-1750$~V. The observation detector PMTs of Hamamatsu Type 1949-50 or 1949-51 were negatively biased using either a CAEN R1470ET or CAEN NDT1470 power supply. 

\begin{figure}
	\centering
	\includegraphics[width=0.47\textwidth]{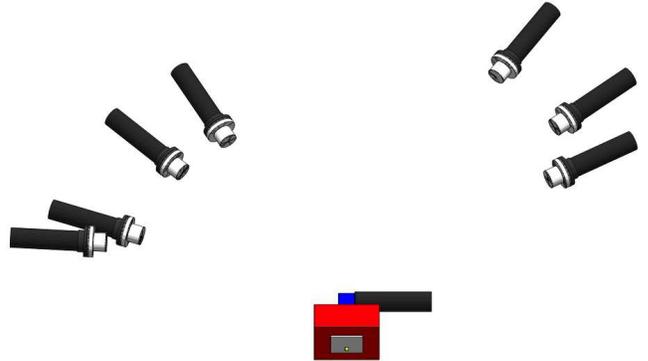}
	\caption{(Color online) Schematic of the experimental setup used to isolate neutron-capture events. The EJ-254 scintillator is represented in blue, the polyethylene shielding in red, and the lead shielding in gray. The AmBe source (gold) was positioned flush against the lead shielding. The target scintillator was atop a sheet of polyethylene and all detectors rested upon a large aluminum table (not pictured).}
	\label{nCaptureSchematic}
\end{figure}

A diagram of the experimental setup is shown in Fig.~\ref{nCaptureSchematic}. The source, positioned behind shielding material, and the target scintillator, resting on a sheet of polyethylene, were placed on a large aluminum table. A minimum of 5~mm of Pb and 5.1~cm of polyethylene were placed between the EJ-254 detector and the source. The Pb shielding reduced contributions from the 59.5~keV $\gamma$ ray produced by decay of $^{241}$Am, and the polyethylene moderator slowed neutrons to increase the neutron-capture rate. For\ $^{10}$B($n,\alpha$) events in the target scintillator populating the first excited state of $^{7}$Li, a 477.6~keV $\gamma$ ray was emitted, which was detected in coincidence with the signal produced in the EJ-254 detector by the recoiling reaction products. Data acquisition and pulse processing were accomplished as described in Section~\ref{daq-dpp}. Data were recorded in two modes: a singles trigger, where events were logged when a signal was observed in the target detector, and a coincident trigger, where events were logged when a signal was observed in the target detector and at least one observation detector within a 160~ns window. The observation detector locations were selected to balance detection efficiency with a sufficient difference in the TOF between fast neutron and $\gamma$-ray interactions to enable rejection of neutron scattering events.

\begin{figure}
	\centering
	\includegraphics[width=0.5\textwidth]{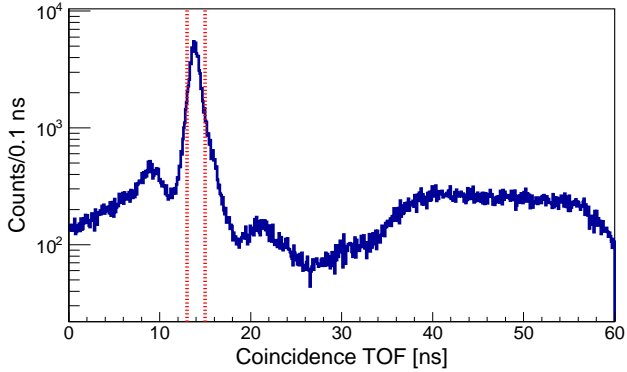}
	\caption{Coincident TOF plot for a single observation detector. The 2~ns window used to select capture-gated events is represented by the vertical dashed lines.} 
	\label{Timing.png}
\end{figure}
 
\subsection{Data Analysis and Calibration}
 
Fig.~\ref{Timing.png} shows the distribution of events in TOF space for a single observation detector. The feature centered at approximately 14~ns corresponds to $\gamma$-$\gamma$ scatters (i.e., $\gamma$-ray scattering in the target detector followed by interaction in the observation detector) and neutron-capture events in the target detector accompanied by the detection of a 477.6~keV $\gamma$ ray in the observation detector. A 2~ns window (indicated by the dashed vertical red lines in Fig.~\ref{Timing.png}) was applied to isolate these events in post-processing.  As described by Sun et al.~\cite{Sun2019}, there is a significant contribution to the overall counts produced by $\gamma$ rays scattering first off an observation detector and then back into the target, visible in coincident TOF space as a smaller peak earlier in time (at $\sim$9~ns). The broad feature (from 35 ns to 60 ns) corresponds to neutron scattering events in the target and observation detectors. Finally, the feature located around 22~ns is the result of $\gamma$-rays scattering off the floor and surrounding materials. 

\begin{figure}
	\centering
	\includegraphics[width=0.5\textwidth]{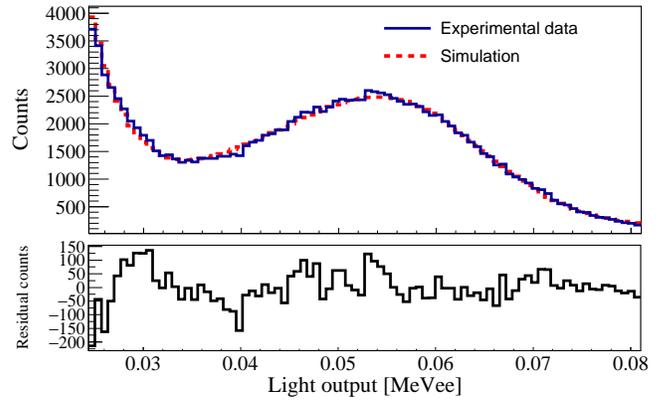}
	\caption{(Color online) The top panel displays the minimization between a simulated energy-deposition spectrum folded with a resolution function (red) and an experimentally-measured pulse integral spectrum (blue) produced with an $^{241}$Am source incident on the EJ-254 detector. The residual plot in the bottom panel demonstrates the goodness of fit.}
	\label{Am241}
\end{figure}

The light output of the EJ-254 target detector was calibrated using the 59.5 keV $\gamma$ ray from an \textsuperscript{241}Am source, placed $>$5 cm from the detector to ensure uniformity of the electron recoil distribution. A Geant4 simulation was used to model the energy deposition in the target, taking into account the surrounding materials, as well as the vinyl electrical tape, aluminum foil, and polytetrafluoroethylene tape used to wrap the scintillator for improved light collection~\cite{geant4}. As EJ-254 and EJ-200 are comprised of the same polymer solvent, the EJ-200 electron light yield nonproportionality curve from Payne et al.\ was normalized at 59.5~keV and used to convert the energy deposition spectrum to light output~\cite{Payne11}. In this case, a Geant4 simulation indicated that approximately one-third of events for which the total energy was deposited were a result of multiple scatters. This simulated light response was convolved with a detector resolution function~\cite{dietze}, and the difference between the resulting distribution and the measured spectrum was then minimized using the SIMPLEX and MIGRAD algorithms from the ROOT Minuit2 package \cite{Brun97}. Fig.~\ref{Am241} shows the light output calibration spectrum, with the residuals calculated as the binwise difference between the measured and simulated spectra. 

Due to the low energy of this $\gamma$ ray and relatively high interaction probability, the calibration was sensitive to the surrounding materials. The amount of light per unit electron-energy deposited extracted using a full simulation of the materials surrounding the detector (accounting for down-scattered $\gamma$ rays) was $8.4\%$ lower than that obtained via simulation assuming a bare scintillator. Neglecting the effect of electron light yield nonproportionality increased the ostensible mean electron-energy equivalent of the distribution of events in the peak region by $3.4\%$.

\subsection{Results and Discussion}

\begin{figure}
	\centering
	\subfloat[\label{PSD-EJ254}]{\includegraphics[width=0.97\columnwidth]{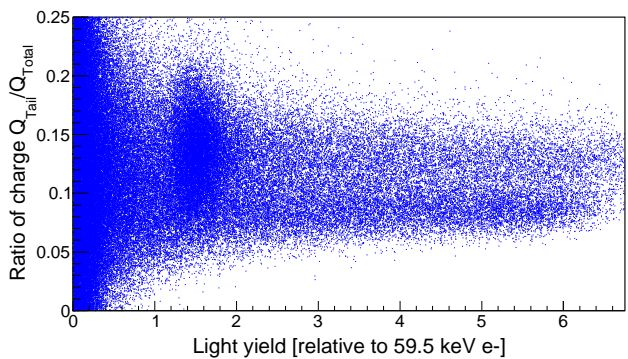}}	
	
	\subfloat[\label{PSDRedandBlack}]{\includegraphics[width=0.97\columnwidth]{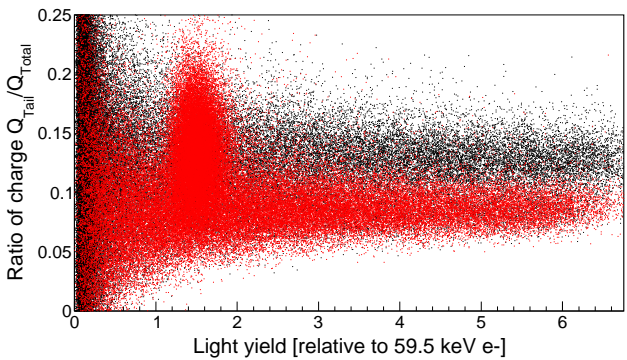}}	
	\caption{(Color online) PSD plot for EJ-254 as a function of the relative light yield. (a) Without coincident constraints. (b) With coincident TOF constraints, where the black data points correspond to neutron scattering events and the red correspond to neutron-capture and $\gamma$-ray scattering events.}
	\label{DualPSDPlot}
\end{figure}

Fig.~\ref{DualPSDPlot} shows the PSD parameter as a function of the total light output in the EJ-254 scintillator. The shape of the scintillation pulse is expected to vary based upon the specific energy loss of the recoil particles. The ratio of prompt-to-delayed light is greater for recoil electrons (arising from $\gamma$-ray interactions) than recoil protons (from fast neutron interactions), and this ratio is even lower for recoiling alpha particles, which dominate the boron neutron-capture scintillation response. Using a charge integration approach, the PSD parameter was defined as the ratio of the integral of the tail of the EJ-254 pulse to that of the total (with a total pulse integration length set to 300~ns). To optimize separation between $\gamma$-ray and fast neutron interactions, the tail start time was sampled over the range of $16-40$~ns in 2~ns increments and the figure-of-merit (i.e., ratio of the peak separation to the sum of the full-width at half-maximum for each particle distribution \cite{Langeveld2017}) was maximized at high relative light yield to obtain a tail start time of 28~ns. Fig.~\ref{PSD-EJ254} shows the PSD plot for the EJ-254 target scintillator obtained without coincident constraints, wherein clear separation between the $\gamma$-ray and neutron distributions was not observed. In Fig.~\ref{PSDRedandBlack}, coincident TOF constraints were applied to separate neutron scattering events (black) and neutron-capture/$\gamma$-scattering events (red). The neutron-capture feature is centered at approximately 1.5 relative light units and overlaps both the $\gamma$-ray and fast neutron bands. 

\begin{figure}
	\centering
	\includegraphics[width=0.5\textwidth]{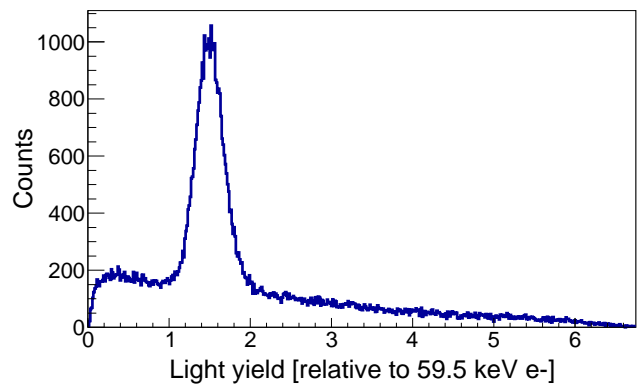}
	\caption{Relative light output spectrum for capture-gated events in the EJ-254 scintillator.}
	\label{ThermalFeatureCounts}
\end{figure}

\begin{table*}[h]
	\caption{Comparison of the relative light yield of neutron-capture events in commercially available boron-loaded organic scintillators.  \label{captureLocComparison}}
	\centering
	\renewcommand{\arraystretch}{1.2}
	\begin{tabular}{ccccc}
		\hline
		Commercial name & Type & Neutron-capture light yield (keVee) & Calibration source & Citation \\ \hline
		EJ-254 & Plastic	& 89.4 $\pm$ 1.1 &$^{241}$Am & This work\\ 
		BC-454 & Plastic & 76 & $^{137}$Cs & Drake et al.~\cite{Drake1986}\\
		EJ-254 & Plastic & 76 & $^{137}$Cs & Eljen~\cite{EJ-254,Drake1986}\\
		EJ-254 & Plastic & 100 &$^{241}$Am & Sun et al.~\cite{Sun2019}\\
		BC-454 & Plastic & 100 & $^{241}$Am & Normand et al.~\cite{Normand2002}\\
		BC-454 & Plastic & 93 & Multiple & Miller et al.~\cite{miller1999neutron}\\
		BC-454 & Plastic & 93 & Multiple & Feldman et al.~\cite{feldman1991novel}\\
		EJ-309B5 & Liquid & 100 &$^{241}$Am & Swiderski et al.~\cite{Swiderski2010} \\
		BC-523 & Liquid & 60 & $^{241}$Am & Aoyama et al.~\cite{Aoyama1993}\\ 
		BC-523A2 & Liquid & 60 & $^{241}$Am & Swiderski et al.~\cite{Swiderski2010}\\
		EJ-339A & Liquid & 60 &$^{241}$Am & Swiderski et al.~\cite{Swiderski2010}\\
		EJ-339A & Liquid & 50 &$^{137}$Cs & Pino et al.~\cite{Pino2014}\\
		\hline
	\end{tabular}
\end{table*}

Fig.~\ref{ThermalFeatureCounts} provides a capture-gated relative light yield histogram for EJ-254, with additional minimum pulse integral constraints applied to the signals in each observation detector to reduce background contributions. The feature was fit using a Gaussian distribution with a power-law background to estimate its centroid value of 1.503 $\pm$ 0.018 relative light units. For the purpose of comparison to the literature, the integrated charge was expressed in units of keVee using the $^{241}$Am photopeak as the calibration fiducial, yielding 89.4 $\pm$ 1.1~keVee. It is important to note that the nonproportionality of the electron response can introduce bias in the reported results if not treated properly. The keVee unit, which is often applied with an incorrect assumption of electron light proportionality, will depend upon the calibration source used unless a correction is applied. The nonproportionality measurement of EJ-200 by Payne et al.~\cite{Payne11} suggests that the light yield (in number of photons per electron energy deposited) is approximately $11\%$ lower for a 59.5~keV electron compared to that for a 477~keV electron (corresponding to the Compton edge of a 662~keV $\gamma$ ray from a $^{137}$Cs source).

To quantify the uncertainty on the neutron-capture light yield, calibration data were recorded before and after the 4-day experiment, and the resulting calibration constants differed by 1.1\%. This discrepancy is likely due to PMT instability during the counting period. This difference was investigated by isolating the neutron-capture feature from 39 chronological segments of the experiment, fitting each feature, and calculating the standard deviation of the resulting centroid values, which exhibited no global trend in time and were instead scattered stochastically about the mean value. This resulted in a $\pm$0.7\% uncertainty on the calibration constant, which was determined as the average of that obtained in the two source measurements. A further 0.4\% uncertainty was observed when varying the fit region around the peak (as a proxy for model error), and was added in quadrature. The result was propagated along with a 0.1\% statistical uncertainty obtained on the fit parameter in the $\chi^2$ minimization. Contributions from the uncertainty in the thickness of aluminum, vinyl tape, and polytetrafluoroethylene wrappings to the calibration constant were investigated using Geant4 simulations and determined to be negligible. A Geant4 simulation was also performed to evaluate the potential for bias in the neutron-capture light yield characterization as a result of 477.6~keV $\gamma$ rays scattering in the volume, and the contribution was determined to be negligible within statistical uncertainty.

Table~\ref{captureLocComparison} presents a comparison of the neutron-capture light yield for commercially available boron-loaded organic scintillators. The measurement presented here, along with those from Sun et al.~\cite{Sun2019} and Miller et al.~\cite{miller1999neutron}, used a coincident tagging setup, whereby 477.6~keV of the energy released in the nuclear reaction is lost to the deexcitation $\gamma$ ray. All other datasets include neutron-capture events directly populating the ground state of $^{7}$Li (with a 6\% thermal branching ratio), leading to higher energy recoiling charged particles. When taking into account electron light nonproportionality, the light yield corresponding to the neutron-capture feature obtained in this work is in rough agreement with previous measurements on EJ-254 and its commercial equivalent, BC-454. Drake et al.\ measured the light output of the capture feature in BC-454 using a $^{137}$Cs source for calibration and obtained a light yield of 76~keVee. Eljen Technology also reports this value in the EJ-254 material specification sheet~\cite{EJ-254}. Correcting for electron light nonproportionality, this number is equivalent to 85.4~keVee.

The measurements in Sun et al.\cite{Sun2019} and Normand et al. \cite{Normand2002} used an $^{241}$Am source for calibration and obtained a higher light output for the neutron capture feature. These discrepancies are likely due to the lack of proper treatment of electron light nonproportionality in the calibration procedures. Neglecting the contribution of down-scattered $\gamma$ rays from materials surrounding the scintillator and electron light yield nonproportionality, the light yield of the neutron capture feature presented in this work would have been estimated at 101.7~keVee. The measurements in~\cite{miller1999neutron, feldman1991novel} used multiple $\gamma$-ray sources to estimate the light yield on an electron-equivalent energy scale. Their results are approximately 4\% higher than the neutron-capture light yield measured in this work. While measurements on commercially available liquids, such as EJ-309B5, show a similar capture-feature light yield to that obtained in~\cite{Sun2019, Normand2002}, a lower relative light yield is reported for BC-523, BC-523A2, and EJ-339A---possibly indicative of higher ionization density and/or oxygen quenching in these materials~\cite{Swiderski2010}.

\section{Summary}
\label{sec:sum}

The fast neutron response of the EJ-254 boron-loaded plastic scintillator was characterized using a double TOF technique for proton recoil energies between approximately 270~keV and 4.5~MeV. The EJ-254 proton light yield agreed with that of EJ-204 within the associated uncertainties and is thus similar to a number of fast plastic scintillators produced by Eljen Technology, including EJ-200, EJ-208, EJ-230, EJ-232, and EJ-232Q \cite{Laplace2020, Manfredi2020}. This is reasonable considering all these materials share the same PVT base and the ionization quenching effect occurs in the host material. The neutron capture response of EJ-254 was measured using a coincident tagging technique. The light output of the neutron-capture feature was determined to be 89.4 $\pm$ 1.1~keVee, or 1.503 $\pm$ 0.018 light units relative to a 59.5~keV electron. This work  showcases the importance of the proper treatment of electron light nonproportionality in scintillator characterization and facilitates the use of EJ-254 in a wide range of applications.

\section*{Appendix}
\label{app-exp}

The relative proton light yield data for EJ-254 are provided in Table~\ref{resultsTable254}.

\begin{table}[!hb]
	\centering
	\renewcommand{\arraystretch}{1.4}
		\caption{Relative proton light yield data for the EJ-254 scintillator. Proton recoil energy bin widths are provided, as well as the light output uncertainties. The covariance matrix is available upon request.}
	\setlength{\tabcolsep}{8pt}
	\begin{tabular}{cc}
		\hline
		Proton recoil energy & Light yield  \\
		 $[$MeV$]$ &  $[$relative to 477~keV electron$]$\\
		\hline
		0.270$_{-0.020}^{+0.020}$ & 0.0546 $\pm$ 0.0016 \\
		0.312$_{-0.023}^{+0.023}$ & 0.0648 $\pm$ 0.0019 \\
		0.362$_{-0.027}^{+0.025}$ & 0.0779 $\pm$ 0.0025 \\ 
		0.418$_{-0.031}^{+0.030}$ & 0.0946 $\pm$ 0.0030 \\ 
		0.483$_{-0.035}^{+0.034}$ & 0.1172 $\pm$ 0.0036 \\ 
		0.557$_{-0.040}^{+0.040}$ & 0.1458 $\pm$ 0.0048 \\ 
		0.643$_{-0.045}^{+0.047}$ & 0.1821 $\pm$ 0.0058 \\ 
		0.742$_{-0.053}^{+0.053}$ & 0.2292 $\pm$ 0.0074 \\ 
		0.856$_{-0.060}^{+0.062}$ & 0.2873 $\pm$ 0.0095 \\ 
		0.986$_{-0.068}^{+0.072}$ & 0.3598 $\pm$ 0.0120 \\ 
		1.138$_{-0.080}^{+0.082}$ & 0.4555 $\pm$ 0.0149 \\ 
		1.312$_{-0.091}^{+0.096}$ & 0.5738 $\pm$ 0.0194 \\ 
		1.514$_{-0.106}^{+0.110}$ & 0.7249 $\pm$ 0.0233 \\ 
		1.612$_{-0.112}^{+0.120}$ & 0.8055 $\pm$ 0.0268 \\ 
		1.862$_{-0.131}^{+0.137}$ & 1.017 $\pm$ 0.0340 \\ 
		2.148$_{-0.150}^{+0.161}$ & 1.285 $\pm$ 0.0441 \\ 
		2.483$_{-0.174}^{+0.187}$ & 1.613 $\pm$ 0.0569 \\ 
		2.868$_{-0.197}^{+0.223}$ & 2.016 $\pm$ 0.0725 \\ 
		3.322$_{-0.231}^{+0.259}$ & 2.534 $\pm$ 0.0931 \\ 
		3.845$_{-0.264}^{+0.301}$ & 3.158 $\pm$ 0.1166 \\
		4.455$_{-0.301}^{+0.369}$ & 3.695 $\pm$ 0.1397 \\ 
		\hline
	\end{tabular}
	\label{resultsTable254}
\end{table}

\section*{Acknowledgements}
The authors thank the 88-Inch Cyclotron operations and facilities staff for their help in performing these experiments.  

\bibliographystyle{IEEEtran}
\bibliography{IEEEabrv,EJ254}

\begin{thebibliography}{10}
\providecommand{\url}[1]{#1}
\csname url@samestyle\endcsname
\providecommand{\newblock}{\relax}
\providecommand{\bibinfo}[2]{#2}
\providecommand{\BIBentrySTDinterwordspacing}{\spaceskip=0pt\relax}
\providecommand{\BIBentryALTinterwordstretchfactor}{4}
\providecommand{\BIBentryALTinterwordspacing}{\spaceskip=\fontdimen2\font plus
\BIBentryALTinterwordstretchfactor\fontdimen3\font minus
  \fontdimen4\font\relax}
\providecommand{\BIBforeignlanguage}[2]{{%
\expandafter\ifx\csname l@#1\endcsname\relax
\typeout{** WARNING: IEEEtran.bst: No hyphenation pattern has been}%
\typeout{** loaded for the language `#1'. Using the pattern for}%
\typeout{** the default language instead.}%
\else
\language=\csname l@#1\endcsname
\fi
#2}}
\providecommand{\BIBdecl}{\relax}
\BIBdecl

\bibitem{Duckworth1950}
\BIBentryALTinterwordspacing
J.~C. Duckworth, A.~W. Merrison, and A.~Whittaker, ``{A High-Efficiency Neutron
  Detector},'' \emph{Nature}, vol. 165, no. 4185, p.~69, 1950. [Online].
  Available: \url{https://doi.org/10.1038/165069a0}
\BIBentrySTDinterwordspacing

\bibitem{Drake1986}
\BIBentryALTinterwordspacing
D.~Drake, W.~Feldman, and C.~Hurlbut, ``New electronically black neutron
  detectors,'' \emph{Nuclear Instruments and Methods in Physics Research
  Section A: Accelerators, Spectrometers, Detectors and Associated Equipment},
  vol. 247, no.~3, pp. 576 -- 582, 1986. [Online]. Available:
  \url{http://www.sciencedirect.com/science/article/pii/0168900286904195}
\BIBentrySTDinterwordspacing

\bibitem{Pawelczak2014}
\BIBentryALTinterwordspacing
{I. A. Pawe\l{}czak and A. M. Glenn and H. P. Martinez and M. L. Carman and N.
  P. Zaitseva and S. A. Payne}, ``Boron-loaded plastic scintillator with
  neutron-$\gamma$ pulse shape discrimination capability,'' \emph{Nuclear
  Instruments and Methods in Physics Research Section A: Accelerators,
  Spectrometers, Detectors and Associated Equipment}, vol. 751, pp. 62 -- 69,
  2014. [Online]. Available:
  \url{http://www.sciencedirect.com/science/article/pii/S0168900214003210}
\BIBentrySTDinterwordspacing

\bibitem{Yemam2015}
H.~A. Yemam, A.~Mahl, U.~Koldemir, T.~Remedes, S.~Parkin, U.~Greife, and
  A.~Sellinger, ``Boron-rich benzene and pyrene derivatives for the detection
  of thermal neutrons,'' \emph{Scientific Reports}, vol.~5, no.~1, p. 1?9, Sep
  2015.

\bibitem{ZaitsevaIAEA}
N.~Zaitseva, A.~Glenn, L.~Carman, A.~Mabe, and S.~Payne, ``New solid-state
  organic scintillators for wide-energy neutron detection,'' Lawrence Livermore
  National Laboratory, Tech. Rep. {LLNL-PROC-742084}, 2017.

\bibitem{Yen2000}
\BIBentryALTinterwordspacing
Y.-F. Yen, J.~Bowman, R.~Bolton, B.~Crawford, P.~Delheij, G.~Hart, T.~Haseyama,
  C.~Frankle, M.~Iinuma, J.~Knudson, A.~Masaike, Y.~Masuda, Y.~Matsuda,
  G.~Mitchell, S.~Penttilä, N.~Roberson, S.~Seestrom, E.~Sharapov, H.~Shimizu,
  D.~Smith, S.~Stephenson, J.~Szymanski, S.~Yoo, and V.~Yuan, ``A high-rate
  $^{10}${B}-loaded liquid scintillation detector for parity-violation studies
  in neutron resonances,'' \emph{Nuclear Instruments and Methods in Physics
  Research Section A: Accelerators, Spectrometers, Detectors and Associated
  Equipment}, vol. 447, no.~3, pp. 476 -- 489, 2000. [Online]. Available:
  \url{http://www.sciencedirect.com/science/article/pii/S016890029901308X}
\BIBentrySTDinterwordspacing

\bibitem{Rasolonjatovo2002}
\BIBentryALTinterwordspacing
A.~Rasolonjatovo, T.~Shiomi, E.~Kim, T.~Nakamura, T.~Nunomiya, A.~Endo,
  Y.~Yamaguchi, and M.~Yoshizawa, ``Development of a new neutron monitor using
  a boron-loaded organic liquid scintillation detector,'' \emph{Nuclear
  Instruments and Methods in Physics Research Section A: Accelerators,
  Spectrometers, Detectors and Associated Equipment}, vol. 492, no.~3, pp. 423
  -- 433, 2002. [Online]. Available:
  \url{http://www.sciencedirect.com/science/article/pii/S0168900202013955}
\BIBentrySTDinterwordspacing

\bibitem{Ishikawa2004}
\BIBentryALTinterwordspacing
M.~Ishikawa, K.~Ono, Y.~Sakurai, H.~Unesaki, A.~Uritani, G.~Bengua,
  T.~Kobayashi, K.~Tanaka, and T.~Kosako, ``Development of real-time thermal
  neutron monitor using boron-loaded plastic scintillator with optical fiber
  for boron neutron capture therapy,'' \emph{Applied Radiation and Isotopes},
  vol.~61, no.~5, pp. 775 -- 779, 2004, topics in Neutron Capture Therapy:
  Proceedings of the Eleventh World Congress on Neutron Capture Therapy
  (ISNCT-11). [Online]. Available:
  \url{http://www.sciencedirect.com/science/article/pii/S0969804304003173}
\BIBentrySTDinterwordspacing

\bibitem{Swiderski2010}
L.~{Swiderski}, M.~{Moszynski}, D.~{Wolski}, T.~{Batsch}, J.~{Iwanowska},
  A.~{Nassalski}, A.~{Syntfeld-Kazuch}, T.~{Szczesniak}, F.~{Kniest}, M.~R.
  {Kusner}, G.~{Pausch}, J.~{Stein}, W.~{Klamra}, P.~{Schotanus}, and
  C.~{Hurlbut}, ``Further study of boron-10 loaded liquid scintillators for
  detection of fast and thermal neutrons,'' \emph{IEEE Transactions on Nuclear
  Science}, vol.~57, no.~1, pp. 375--380, Feb 2010.

\bibitem{Pino2014}
\BIBentryALTinterwordspacing
F.~Pino, L.~Stevanato, D.~Cester, G.~Nebbia, L.~Sajo-Bohus, and G.~Viesti,
  ``{Detecting fast and thermal neutrons with a boron loaded liquid
  scintillator, {EJ-339A}},'' \emph{Applied Radiation and Isotopes}, vol.~92,
  pp. 6 -- 11, 2014. [Online]. Available:
  \url{http://www.sciencedirect.com/science/article/pii/S0969804314002280}
\BIBentrySTDinterwordspacing

\bibitem{Agnes2016}
\BIBentryALTinterwordspacing
P.~Agnes, L.~Agostino, I.~Albuquerque, T.~Alexander, A.~Alton, K.~Arisaka,
  H.~Back, B.~Baldin, K.~Biery, G.~Bonfini, M.~Bossa, B.~Bottino, A.~Brigatti,
  J.~Brodsky, F.~Budano, S.~Bussino, M.~Cadeddu, L.~Cadonati, M.~Cadoni,
  F.~Calaprice, N.~Canci, A.~Candela, H.~Cao, M.~Cariello, M.~Carlini,
  S.~Catalanotti, P.~Cavalcante, A.~Chepurnov, A.~Cocco, G.~Covone, L.~Crippa,
  D.~D'Angelo, M.~D'Incecco, S.~Davini, S.~D. Cecco, M.~D. Deo, M.~D. Vincenzi,
  A.~Derbin, A.~Devoto, F.~D. Eusanio, G.~D. Pietro, E.~Edkins, A.~Empl,
  A.~Fan, G.~Fiorillo, K.~Fomenko, G.~Foster, D.~Franco, F.~Gabriele,
  C.~Galbiati, C.~Giganti, A.~Goretti, F.~Granato, L.~Grandi, M.~Gromov,
  M.~Guan, Y.~Guardincerri, B.~Hackett, K.~Herner, E.~Hungerford, A.~Ianni,
  A.~Ianni, I.~James, T.~Johnson, C.~Jollet, K.~Keeter, C.~Kendziora,
  V.~Kobychev, G.~Koh, D.~Korablev, G.~Korga, A.~Kubankin, X.~Li, M.~Lissia,
  P.~Lombardi, S.~Luitz, Y.~Ma, I.~Machulin, A.~Mandarano, S.~Mari, J.~Maricic,
  L.~Marini, C.~Martoff, A.~Meregaglia, P.~Meyers, T.~Miletic, R.~Milincic,
  D.~Montanari, A.~Monte, M.~Montuschi, M.~Monzani, P.~Mosteiro, B.~Mount,
  V.~Muratova, P.~Musico, J.~Napolitano, A.~Nelson, S.~Odrowski, M.~Orsini,
  F.~Ortica, L.~Pagani, M.~Pallavicini, E.~Pantic, S.~Parmeggiano, K.~Pelczar,
  N.~Pelliccia, S.~Perasso, A.~Pocar, S.~Pordes, D.~Pugachev, H.~Qian,
  K.~Randle, G.~Ranucci, A.~Razeto, B.~Reinhold, A.~Renshaw, A.~Romani,
  B.~Rossi, N.~Rossi, S.~Rountree, D.~Sablone, P.~Saggese, R.~Saldanha,
  W.~Sands, S.~Sangiorgio, C.~Savarese, E.~Segreto, D.~Semenov, E.~Shields,
  P.~Singh, M.~Skorokhvatov, O.~Smirnov, A.~Sotnikov, C.~Stanford, Y.~Suvorov,
  R.~Tartaglia, J.~Tatarowicz, G.~Testera, A.~Tonazzo, P.~Trinchese,
  E.~Unzhakov, A.~Vishneva, R.~Vogelaar, M.~Wada, S.~Walker, H.~Wang, Y.~Wang,
  A.~Watson, S.~Westerdale, J.~Wilhelmi, M.~Wojcik, X.~Xiang, J.~Xu, C.~Yang,
  J.~Yoo, S.~Zavatarelli, A.~Zec, W.~Zhong, C.~Zhu, and G.~Zuzel, ``The veto
  system of the {DarkSide}-50 experiment,'' \emph{Journal of Instrumentation},
  vol.~11, no.~03, pp. P03\,016--P03\,016, mar 2016. [Online]. Available:
  \url{https://doi.org/10.1088/1748-0221/11/03/P03016}
\BIBentrySTDinterwordspacing

\bibitem{Westerdale2017}
\BIBentryALTinterwordspacing
S.~Westerdale, J.~Xu, E.~Shields, F.~Froborg, F.~Calaprice, T.~Alexander,
  A.~Aprahamian, H.~Back, C.~Casarella, X.~Fang, Y.~Gupta, E.~Lamere, Q.~Liu,
  S.~Lyons, M.~Smith, and W.~Tan, ``Quenching measurements and modeling of a
  boron-loaded organic liquid scintillator,'' \emph{Journal of
  Instrumentation}, vol.~12, no.~08, pp. P08\,002--P08\,002, aug 2017.
  [Online]. Available: \url{https://doi.org/10.1088/1748-0221/12/08/P08002}
\BIBentrySTDinterwordspacing

\bibitem{Peerani}
\BIBentryALTinterwordspacing
P.~Peerani, A.~Tomanin, S.~Pozzi, J.~Dolan, E.~Miller, M.~Flaska,
  M.~Battaglieri, R.~D. Vita], L.~Ficini, G.~Ottonello, G.~Ricco, G.~Dermody,
  and C.~Giles, ``Testing on novel neutron detectors as alternative to
  $^{3}${He} for security applications,'' \emph{Nuclear Instruments and Methods
  in Physics Research Section A: Accelerators, Spectrometers, Detectors and
  Associated Equipment}, vol. 696, pp. 110 -- 120, 2012. [Online]. Available:
  \url{http://www.sciencedirect.com/science/article/pii/S0168900212007929}
\BIBentrySTDinterwordspacing

\bibitem{Swiderski2008}
L.~{Swiderski}, M.~{Moszynski}, D.~{Wolski}, T.~{Batsch}, A.~{Nassalski},
  A.~{Syntfeld-Kazuch}, T.~{Szczesniak}, F.~{Kniest}, M.~R. {Kusner},
  G.~{Pausch}, J.~{Stein}, and W.~{Klamra}, ``Boron-10 loaded {BC523A} liquid
  scintillator for neutron detection in the border monitoring,'' \emph{IEEE
  Transactions on Nuclear Science}, vol.~55, no.~6, pp. 3710--3716, Dec 2008.

\bibitem{Dazeley2018}
S.~Dazeley, A.~Bernstein, T.~Classen, E.~Reedy, D.~Hellfeld, M.~Duvall, and
  C.~Marianno, ``Antineutrino detection based on $^{6}${Li}-doped pulse shape
  sensitive plastic scintillator and gadolinium-doped water,''
  \emph{International Journal of Modern Physics: Conference Series}, vol.~48,
  p. 1860105, 2018.

\bibitem{EJ-254}
\BIBentryALTinterwordspacing
\emph{{Boron Loaded Plastic Scintillator EJ-254}}, Eljen Technology, July 2020.
  [Online]. Available:
  \url{https://eljentechnology.com/images/products/data_sheets/EJ-254.pdf}
\BIBentrySTDinterwordspacing

\bibitem{Greenwood1979}
L.~R. Greenwood and N.~R. Chellew, ``Improved $^{10}$b-loaded liquid
  scintillator with pulse-shape discrimination,'' \emph{Review of Scientific
  Instruments}, vol.~50, no.~4, pp. 466--471, 1979.

\bibitem{Klein2007}
H.~Klein and F.~D. Brooks, ``{Scintillation Detectors For Fast Neutrons},'' in
  \emph{Proceedings of International Workshop on Fast Neutron Detectors and
  Applications {\textemdash} PoS(FNDA2006)}, vol. 025, 2007, p. 097.

\bibitem{Aoyama1993}
\BIBentryALTinterwordspacing
T.~Aoyama, K.~Honda, C.~Mori, K.~Kudo, and N.~Takeda, ``Energy response of a
  full-energy-absorption neutron spectrometer using boron-loaded liquid
  scintillator {BC-523},'' \emph{Nuclear Instruments and Methods in Physics
  Research Section A: Accelerators, Spectrometers, Detectors and Associated
  Equipment}, vol. 333, no.~2, pp. 492 -- 501, 1993. [Online]. Available:
  \url{http://www.sciencedirect.com/science/article/pii/016890029391197U}
\BIBentrySTDinterwordspacing

\bibitem{Weinfurther2018}
\BIBentryALTinterwordspacing
K.~Weinfurther, J.~Mattingly, E.~Brubaker, and J.~Steele, ``Model-based design
  evaluation of a compact, high-efficiency neutron scatter camera,''
  \emph{Nuclear Instruments and Methods in Physics Research Section A:
  Accelerators, Spectrometers, Detectors and Associated Equipment}, vol. 883,
  pp. 115 -- 135, 2018. [Online]. Available:
  \url{http://www.sciencedirect.com/science/article/pii/S0168900217312238}
\BIBentrySTDinterwordspacing

\bibitem{Brown2018}
D.~Brown, M.~Chadwick, R.~Capote, A.~Kahler, A.~Trkov, M.~Herman, A.~Sonzogni,
  Y.~Danon, A.~Carlson, M.~Dunn, D.~Smith, G.~Hale, G.~Arbanas, R.~Arcilla,
  C.~Bates, B.~Beck, B.~Becker, F.~Brown, R.~Casperson, J.~Conlin, D.~Cullen,
  M.-A. Descalle, R.~Firestone, T.~Gaines, K.~Guber, A.~Hawari, J.~Holmes,
  T.~Johnson, T.~Kawano, B.~Kiedrowski, A.~Koning, S.~Kopecky, L.~Leal,
  J.~Lestone, C.~Lubitz, J.~{Márquez Damián}, C.~Mattoon, E.~McCutchan,
  S.~Mughabghab, P.~Navratil, D.~Neudecker, G.~Nobre, G.~Noguere, M.~Paris,
  M.~Pigni, A.~Plompen, B.~Pritychenko, V.~Pronyaev, D.~Roubtsov, D.~Rochman,
  P.~Romano, P.~Schillebeeckx, S.~Simakov, M.~Sin, I.~Sirakov, B.~Sleaford,
  V.~Sobes, E.~Soukhovitskii, I.~Stetcu, P.~Talou, I.~Thompson, S.~{van der
  Marck}, L.~Welser-Sherrill, D.~Wiarda, M.~White, J.~Wormald, R.~Wright,
  M.~Zerkle, G.~Žerovnik, and Y.~Zhu, ``{ENDF/B-VIII.0: T}he 8th major release
  of the nuclear reaction data library with {CIELO}-project cross sections, new
  standards and thermal scattering data,'' \emph{Nuclear Data Sheets}, vol.
  148, pp. 1 -- 142, 2018, special Issue on Nuclear Reaction Data.

\bibitem{Brown18}
\BIBentryALTinterwordspacing
J.~A. Brown, B.~L. Goldblum, T.~A. Laplace, K.~P. Harrig, L.~A. Bernstein,
  D.~L. Bleuel, W.~Younes, D.~Reyna, E.~Brubaker, and P.~Marleau, ``Proton
  light yield in organic scintillators using a double time-of-flight
  technique,'' \emph{Journal of Applied Physics}, vol. 124, no.~4, p. 045101,
  2018. [Online]. Available: \url{https://doi.org/10.1063/1.5039632}
\BIBentrySTDinterwordspacing

\bibitem{Harrig18}
K.~P. Harrig, B.~L. Goldblum, J.~A. Brown, D.~L. Bleuel, L.~A. Bernstein,
  J.~Bevins, M.~Harasty, T.~A. Laplace, and E.~F. Matthews, ``Neutron
  spectroscopy for pulsed beams with frame overlap using a double
  time-of-flight technique,'' \emph{Nuclear Instruments and Methods in Physics
  Research Section A: Accelerators, Spectrometers, Detectors and Associated
  Equipment}, vol. 877, pp. 359 -- 366, 2018.

\bibitem{Janecek2012}
M.~{Janecek}, ``Reflectivity spectra for commonly used reflectors,'' \emph{IEEE
  Transactions on Nuclear Science}, vol.~59, no.~3, pp. 490--497, June 2012.

\bibitem{Manfredi2020}
J.~J. Manfredi, B.~L. Goldblum, T.~A. Laplace, G.~Gabella, J.~Gordon,
  A.~O'Brien, S.~Chowdhury, J.~A. Brown, and E.~M. Brubaker, ``Proton light
  yield of fast plastic scintillators for neutron imaging,'' \emph{IEEE
  Transactions on Nuclear Science}, vol.~67, pp. 434--442, 2020.

\bibitem{Brun1997}
\BIBentryALTinterwordspacing
R.~Brun and F.~Rademakers, ``{ROOT} -- {An} object oriented data analysis
  framework,'' \emph{Nuclear Instruments and Methods in Physics Research
  Section A: Accelerators, Spectrometers, Detectors and Associated Equipment},
  vol. 389, no.~1, pp. 81 -- 86, 1997. [Online]. Available:
  \url{http://www.sciencedirect.com/science/article/pii/S016890029700048X}
\BIBentrySTDinterwordspacing

\bibitem{Friend11}
M.~Friend, G.~Franklin, and B.~Quinn, ``An {LED} pulser for measuring
  photomultiplier linearity,'' \emph{Nuclear Instruments and Methods in Physics
  Research Section A: Accelerators, Spectrometers, Detectors and Associated
  Equipment}, vol. 676, pp. 66 -- 69, 2012.

\bibitem{Laplace2020}
\BIBentryALTinterwordspacing
T.~A. Laplace, B.~L. Goldblum, J.~A. Brown, D.~L. Bleuel, C.~A. Brand,
  G.~Gabella, T.~Jordan, C.~Moore, N.~Munshi, Z.~W. Sweger, A.~Sweet, and
  E.~Brubaker, ``Low energy light yield of fast plastic scintillators,''
  \emph{Nuclear Instruments and Methods in Physics Research Section A:
  Accelerators, Spectrometers, Detectors and Associated Equipment}, vol. 954,
  p. 161444, 2020, {Symposium on Radiation Measurements and Applications XVII}.
  [Online]. Available:
  \url{http://www.sciencedirect.com/science/article/pii/S0168900218314360}
\BIBentrySTDinterwordspacing

\bibitem{Nassalski08}
A.~Nassalski, M.~Moszy\'{n}ski, A.~Syntfeld-Ka\.{z}uch, L.~\'{S}widerski, and
  T.~Szcz\c{e}\'{s}niak, ``{Non-Proportionality of Organic Scintillators and
  BGO},'' \emph{IEEE Transactions on Nuclear Science}, vol.~55, no.~3, pp.
  1069--1072, June 2008.

\bibitem{Payne11}
S.~A. Payne, W.~W. Moses, S.~Sheets, L.~Ahle, N.~J. Cherepy, B.~Sturm,
  S.~Dazeley, G.~Bizarri, and W.~S. Choong, ``{Nonproportionality of
  Scintillator Detectors: Theory and Experiment. II},'' \emph{IEEE Transactions
  on Nuclear Science}, vol.~58, no.~6, pp. 3392--3402, Dec 2011.

\bibitem{geant4}
S.~Agostinelli, J.~Allison, K.~Amako, J.~Apostolakis, H.~Araujo, P.~Arce,
  M.~Asai, D.~Axen, S.~Banerjee, G.~Barrand, F.~Behner, L.~Bellagamba,
  J.~Boudreau, L.~Broglia, A.~Brunengo, H.~Burkhardt, S.~Chauvie, J.~Chuma,
  R.~Chytracek, G.~Cooperman, G.~Cosmo, P.~Degtyarenko, A.~Dell'Acqua,
  G.~Depaola, D.~Dietrich, R.~Enami, A.~Feliciello, C.~Ferguson, H.~Fesefeldt,
  G.~Folger, F.~Foppiano, A.~Forti, S.~Garelli, S.~Giani, R.~Giannitrapani,
  D.~Gibin, J.~G. Cadenas, I.~Gonzlez, G.~G. Abril, G.~Greeniaus, W.~Greiner,
  V.~Grichine, A.~Grossheim, S.~Guatelli, P.~Gumplinger, R.~Hamatsu,
  K.~Hashimoto, H.~Hasui, A.~Heikkinen, A.~Howard, V.~Ivanchenko, A.~Johnson,
  F.~Jones, J.~Kallenbach, N.~Kanaya, M.~Kawabata, Y.~Kawabata, M.~Kawaguti,
  S.~Kelner, P.~Kent, A.~Kimura, T.~Kodama, R.~Kokoulin, M.~Kossov,
  H.~Kurashige, E.~Lamanna, T.~Lampn, V.~Lara, V.~Lefebure, F.~Lei, M.~Liendl,
  W.~Lockman, F.~Longo, S.~Magni, M.~Maire, E.~Medernach, K.~Minamimoto, P.~M.
  de~Freitas, Y.~Morita, K.~Murakami, M.~Nagamatu, R.~Nartallo, P.~Nieminen,
  T.~Nishimura, K.~Ohtsubo, M.~Okamura, S.~O'Neale, Y.~Oohata, K.~Paech,
  J.~Perl, A.~Pfeiffer, M.~Pia, F.~Ranjard, A.~Rybin, S.~Sadilov, E.~D. Salvo,
  G.~Santin, T.~Sasaki, N.~Savvas, Y.~Sawada, S.~Scherer, S.~Sei, V.~Sirotenko,
  D.~Smith, N.~Starkov, H.~Stoecker, J.~Sulkimo, M.~Takahata, S.~Tanaka,
  E.~Tcherniaev, E.~S. Tehrani, M.~Tropeano, P.~Truscott, H.~Uno, L.~Urban,
  P.~Urban, M.~Verderi, A.~Walkden, W.~Wander, H.~Weber, J.~Wellisch,
  T.~Wenaus, D.~Williams, D.~Wright, T.~Yamada, H.~Yoshida, and D.~Zschiesche,
  ``{Geant4} -- a simulation toolkit,'' \emph{Nuclear Instruments and Methods
  in Physics Research Section A: Accelerators, Spectrometers, Detectors and
  Associated Equipment}, vol. 506, no.~3, pp. 250 -- 303, 2003.

\bibitem{dietze}
G.~Dietze and H.~Klein, ``Gamma-calibration of {NE 213} scintillation
  counters,'' \emph{Nuclear Instruments and Methods in Physics Research}, vol.
  193, no.~3, pp. 549 -- 556, 1982.

\bibitem{Brun97}
R.~Brun and F.~Rademakers, ``{ROOT} -- {An} object oriented data analysis
  framework,'' \emph{Nuclear Instruments and Methods in Physics Research
  Section A: Accelerators, Spectrometers, Detectors and Associated Equipment},
  vol. 389, no.~1, pp. 81 -- 86, 1997.

\bibitem{birks}
J.~Birks, \emph{The Theory and Practice of Scintillation Counting}.\hskip 1em
  plus 0.5em minus 0.4em\relax Pergamon Press, 1964, ch.~1, pp. 447--450.

\bibitem{LaplaceJINST}
\BIBentryALTinterwordspacing
T.~A. Laplace, B.~L. Goldblum, J.~E. Bevins, D.~L. Bleuel, E.~Bourret, J.~A.
  Brown, E.~J. Callaghan, J.~S. Carlson, P.~L. Feng, G.~Gabella, K.~P. Harrig,
  J.~J. Manfredi, C.~Moore, F.~Moretti, M.~Shinner, A.~Sweet, and Z.~W. Sweger,
  ``Comparative scintillation performance of {EJ-309, EJ-276,} and a novel
  organic glass,'' \emph{Journal of Instrumentation}, vol.~15, no.~11, pp.
  P11\,020--P11\,020, nov 2020. [Online]. Available:
  \url{https://doi.org/10.1088%2F1748-0221%2F15%2F11%2Fp11020}
\BIBentrySTDinterwordspacing

\bibitem{Birks1951}
J.~B. Birks, ``Scintillations from organic crystals: Specific fluorescence and
  relative response to different radiations,'' \emph{Proceedings of the
  Physical Society. Section A}, vol.~64, no.~10, pp. 874--877, oct 1951.

\bibitem{birks321}
J.~Birks, \emph{The Theory and Practice of Scintillation Counting}.\hskip 1em
  plus 0.5em minus 0.4em\relax Pergamon Press, 1964, ch.~9, p. 321.

\bibitem{Zaitseva2012}
\BIBentryALTinterwordspacing
N.~Zaitseva, B.~L. Rupert, I.~Pawe\l{}czak, A.~Glenn, H.~P. Martinez,
  L.~Carman, M.~Faust, N.~Cherepy, and S.~Payne, ``Plastic scintillators with
  efficient neutron/gamma pulse shape discrimination,'' \emph{Nuclear
  Instruments and Methods in Physics Research Section A: Accelerators,
  Spectrometers, Detectors and Associated Equipment}, vol. 668, pp. 88 -- 93,
  2012. [Online]. Available:
  \url{http://www.sciencedirect.com/science/article/pii/S0168900211021395}
\BIBentrySTDinterwordspacing

\bibitem{Sun2019}
\BIBentryALTinterwordspacing
Y.~Sun, H.~Zhang, X.~Zhao, M.~Shao, Z.~Tang, and C.~Li, ``Identifying thermal
  neutrons, fast neutrons, and gamma rays by using a scintillator-based
  time-of-flight method,'' \emph{Nuclear Instruments and Methods in Physics
  Research Section A: Accelerators, Spectrometers, Detectors and Associated
  Equipment}, vol. 940, pp. 129 -- 134, 2019. [Online]. Available:
  \url{http://www.sciencedirect.com/science/article/pii/S0168900219308678}
\BIBentrySTDinterwordspacing

\bibitem{Langeveld2017}
W.~G.~J. {Langeveld}, M.~J. {King}, J.~{Kwong}, and D.~T. {Wakeford}, ``Pulse
  shape discrimination algorithms, figures of merit, and gamma-rejection for
  liquid and solid scintillators,'' \emph{IEEE Transactions on Nuclear
  Science}, vol.~64, no.~7, pp. 1801--1809, 2017.

\bibitem{Normand2002}
S.~{Normand}, B.~{Mouanda}, S.~{Haan}, and M.~{Louvel}, ``Study of a new boron
  loaded plastic scintillator (revised),'' \emph{IEEE Transactions on Nuclear
  Science}, vol.~49, no.~4, pp. 1603--1608, Aug 2002.

\bibitem{miller1999neutron}
M.~C. Miller, R.~S. Biddle, S.~C. Bourret, R.~C. Byrd, N.~Ensslin, W.~C.
  Feldman, J.~J. Kuropatwinski, J.~L. Longmire, M.~S. Krick, D.~R. Mayo
  \emph{et~al.}, ``Neutron detection and applications using a {BC454/BGO}
  array,'' \emph{Nuclear Instruments and Methods in Physics Research Section A:
  Accelerators, Spectrometers, Detectors and Associated Equipment}, vol. 422,
  no. 1-3, pp. 89--94, 1999.

\bibitem{feldman1991novel}
W.~C. Feldman, G.~F. Auchampaugh, and R.~C. Byrd, ``A novel fast-neutron
  detector for space applications,'' \emph{Nuclear Instruments and Methods in
  Physics Research Section A: Accelerators, Spectrometers, Detectors and
  Associated Equipment}, vol. 306, no. 1-2, pp. 350--365, 1991.

\end{thebibliography}

\end{document}